\documentclass[12pt]{article}
\title{ New spectral representation and evaluation of  $f_\pi$ and the quark condensate
 $\lan \bar q q\ran$ in the terms of string tension}
\author{Yu.A.Simonov\\
 State Research
Center\\Institute of Theoretical and Experimental Physics, \\
Moscow, 117218 Russia}
 \date{}
\newcommand{\beq}{\begin{eqnarray}}
 \newcommand{\eeq}{\end{eqnarray}}
\newcommand{\be}{\begin{equation}}
 \newcommand{\ee}{\end{equation}}

\def\ga{\mathrel{\mathpalette\fun >}}
\def\fun#1#2{\lower3.6pt\vbox{\baselineskip0pt\lineskip.9pt
\ialign{$\mathsurround=0pt#1\hfil ##\hfil$\crcr#2\crcr\sim\crcr}}}
\newcommand{\veX}{\mbox{\boldmath${\rm X}$}}
\newcommand{{\SD}}{\rm SD}

\newcommand{\vex}{\mbox{\boldmath${\rm x}$}}
\newcommand{\vey}{\mbox{\boldmath${\rm y}$}}
\newcommand{\ver}{\mbox{\boldmath${\rm r}$}}

\newcommand{\veP}{\mbox{\boldmath${\rm P}$}}
\newcommand{\vep}{\mbox{\boldmath${\rm p}$}}

\newcommand{\veR}{\mbox{\boldmath${\rm R}$}}

\newcommand{\vek}{\mbox{\boldmath${\rm k}$}}

\newcommand{\veal}{\mbox{\boldmath${\rm \alpha}$}}

\newcommand{\lan}{\langle}
\newcommand{\ran}{\rangle}
\begin{document}
\maketitle

\begin{abstract}
 New spectral representations for  $f_\pi$
 and chiral condensate are derived in QCD and used  for calculations in the large $N_c$ limit.
  Both quantities are expressed in this limit
through string tension $\sigma$ and gluon correlation length $T_g$
without fitting parameters. As a result one obtains $\lan \bar q
q\ran = - N_c\sigma^2 T_g a_1,~~ f_\pi =\sqrt{N_c} \sigma T_g
a_2$, with $a_1=0.0823, a_2=0.30$. Taking  $\sigma=0.18 $GeV$^2$
and $T_g=1$ GeV$^{-1}$,  as known from analytic and lattice
calculations, this yields $\lan \bar q q\ran (\mu=2 {\rm
GeV})=-(0.225 {\rm GeV})^3$, $ f_\pi=0.094$ GeV, which  is close
to the standard values.

\end{abstract}

\section{ Introduction}

The Chiral Symmetry Breaking (CSB) is known to occur in QCD at
large $N_c$, if confinement is preserved in this limit \cite{1}.
Lattice calculations for $N_c=2,3$ indicate that confinement and
CSB coexist in the confinement phase at $T\leq T_c$ and disappear
simultaneously above $T_c$ \cite{2}. At larger $N_c$ it was found
on the lattice that the $1/N_c$ corrections to all observables
studied are not large \cite{3}, suggesting that a smooth limit at
large $N_c$ is possible.

In the framework of the Field Correlator Method FCM \cite{4} the
dynamics of confinement and deconfinement is associated with the
set of field correlators $D^{(n)}_{\mu_1\nu_1,...\mu_n\nu_n}
(x_1,...x_n)= \lan F_{\mu_1\nu_1} (x_1)... F_{\mu_n\nu_n} (x_n)
\ran$\footnote{parallel transporters are here omitted for
simplicity} of which the lowest one $D^{(2)} (x_1, x_2) \equiv
D^{(2)}(x_1-x_2)$ plays the dominant role \cite{5}. Moreover,
$D^{(2)}(x)$ was calculated on the lattice \cite{6} and  its
confining part, $D(x)$, was shown to disappear exactly above $T_c$
\cite{7}.

In \cite{8,9} also CSB was found as a consequence of confinement
and  in \cite{9,10} the Effective Chiral Lagrangian (ECL) was
derived from the $4q$ interaction term using $D(x)$ as a kernel.

The resulting ECL in \cite{9,10} has a general structure  which
can be reduced to the expressions derived in the framework of the
instanton model \cite{11} or the NJL model \cite{12}, when the
corresponding kernels are introduced there.

In the case of confinement, the effective quark mass operator
$M(x)$ in QCD obtained in \cite{9,10}  contains the effect of the
scalar confining string connecting the quark to the nearest
antiquark. Moreover all invariant quark Green's functions can be
expressed at large $N_c$ through the  string spectrum as it was
done in \cite{10} in the PS channel.

The phenomenon of CSB was shown in \cite{9,10}  as occurring due
to the spontaneous creation of the  scalar string  (similar to the
creation of  the scalar condensate in nonconfining models
\cite{12,13}) which generates CSB and chiral Nambu-Goldstone (NG)
fields (see eqs. (50-54) in \cite{9} and eqs. (21-24) in
\cite{10}).

Since confinement is present in our formalism (in the form of
$M(x)$) one can ask the question how confinement fits in the
chiral picture of NG spectrum, and in particular how CSB modifies
the lowest PS states computed in FCM (or in any quark model)
taking into account confinement and disregarding CSB. Two such
lowest states, $\pi^{(0)}$ and its first radial excitation
$\pi^{(1)}$ with masses $m(\pi^{(0)})\equiv m_0\cong 0.4$ GeV and
$m(\pi^{(1)})\equiv m_1 \cong 1.35 $ GeV have been computed in
FCM, see Appendix 2 below in this paper. It was shown in \cite{10}
that the ECL obtained there with account of confinement, has a
remarkable property: the PS spectrum of confinement transforms due
to CSB in such a way that $\pi^{(0)}$ becomes a NG pion with the
mass satisfying Gell-Mann-Oakes--Renner (GOR) relation \cite{13}
while the first radial excitation shifts only slightly.

In deriving that property it was essential that all basic
quantities in the ECL and in particular the pion self-energy
operator can be expressed as a spectral decomposition in the
confinement (string-like) spectrum states, which is possible in
the large $N_c$ limit.

In this paper we follow this line to obtain a more fundamental
relation, namely, to  calculate the quark condensate $\lan \bar q
q\ran$ and  the pion decay  constant $f_\pi$ using new spectral
representations for these quantities. Since in the latter all
masses and coupling constants are expressed via $D(x)$, i.e. via
the string tension $\sigma$ and the gluon correlation length
$T_g$, we have an  expression for  $\lan \bar q q\ran$ and $f_\pi$
in terms of $\sigma$ and $T_g$. The most important role in the
spectral representations is played by the lowest PS meson
$\pi^{(0)}$  -- the "to be pion" --  which is the quark model
analog of the pion with mass $m_{0}$ shifted by the hyperfine
interaction from the $\rho$-meson mass. In Appendix 2 we derive
the mass $m_{0}$ and the corresponding wave-function in the
framework of FCM in terms of $\sigma$ and $\alpha_s$.

Having established the connection of $\lan \bar q q\ran$ and
$f_\pi$ with $\sigma, T_g$ stated in abstract above, and explained
in the text below, it is easy to understand that at the the
deconfinement transition when $\sigma$ vanishes at $T=T_c$, also
$\lan \bar q q\ran$ and $f_\pi$ vanish in agreement with lattice
data \cite{2}.

Some specification with respect to  the  notion of "magnetic
confinement" \cite{14} is needed at this point  since magnetic
counterpart of $D(x)$ and the corresponding spacial string tension
stay nonzero above $T_c$. This topic  will be studied elsewhere.

The paper is organized as follows. In the next section the ECL is
written down together with the appropriate expressions for
$\lan\bar q q \ran$ and $f_\pi$. In section 3 the spectral
representations for these quantities are derived, with
coefficients depending on eigenfunctions of the $\bar q q$ system
in the pseudoscalar channel.
 Section 4
is devoted to  the discussion of results in comparison to lattice
data  and to the concluding remarks. Four appendices are included
in the paper, containing  respectively the evaluation of $M(0) $,
derivation of spectral representation, explicit calculation of
eigenvalues and eigenfunctions in the pseudoscalar spectrum, and
the contribution of the small-distance region.

\section{The Effective Chiral Lagrangian}

The  quadratic part of the ECL for pions was derived in \cite{10}
and has the form
\be
W^{(2)} (\phi) =\frac{N_c}{2} \int \phi_a (k) \phi_a (-k) \bar
N(k) \frac{d^{(4)}k}{(2\pi)^4}\label{1}\ee where notations of
\cite{10} have been used, $\phi_a=\frac{2\pi_a}{f_\pi}$ and
\be
\bar N(k) =\frac12 [G^{(MM)} (k) + tr (\Lambda M_S)] = (m^2_\pi +
k^2 ) \frac{f^2_\pi}{4N_c} +O(k^4),\label{2}\ee
\be
G^{(MM)} (k) \equiv - \int tr ( \Lambda (y,x){\gamma_5} M_S(x)
\Lambda{(x,y)}{\gamma_5} M_S (y)) e^{ik(x-y)} d^{(4)}
(x-y),\label{3}\ee
\be
\Lambda(x,y) = (\hat \partial +m+M_S)^{-1}_{x,y}.\label{4}\ee

As it was shown in \cite{10}, two terms in the square brackets in
(\ref{2}) cancel for $k^2=m=0$ and one obtains the GOR relation
for the pion mass \cite{13} \be mN_c tr \Lambda \equiv m|\lan \bar
\psi \psi\ran_M| =\frac12 (m_u+m_d)|\lan \bar u u +\bar d d \ran
|= m^2_\pi f^2_\pi.\label{5} \ee

To calculate the quark condensate, defined in the Minkowskian
space time, one can write $\lan \bar \psi \psi \ran_M=-N_c tr
\Lambda$, and use identical transformation $$ tr \Lambda_{xx} =tr
\lan \frac{1}{(M_S+m+\hat \partial)}(M_S+m-\hat \partial )
\frac{1}{(M_S+m-\hat \partial)}\ran=$$
\be
=\int \lan tr (\gamma_5 \Lambda (x,y) \gamma_5 (M_S+m) \Lambda
(y,x)) \ran  d^4y \equiv - \int G^{(M)} (x,y) d^4y \equiv - G^{(M)}
(k=0). \label{6}\ee Hence $tr\Lambda_{xx}$ reduces to the
zero-momentum component of the $q\bar q$ Green's function in the
PS channel, which differs from (\ref{3}) only by vertex operators.

To define $f_\pi$,  one needs the first term in the $k^2$
expansion of $G^{(MM)}(k)$ Eq.(\ref{3}), (c.f. Eq.(\ref{2})) so
that one has
\be
G^{(MM)} (k) - G^{(MM)} (0) = \frac{k^2 f^2_\pi}{2N_c}
+O(k^4)\label{7}\ee As it was argued in \cite{10}, both $G^{(MM)}(k)$
and $G^{(M)}(k)$ have spectral representations in the
large $N_c$ limit, with the same set of poles $m_n, n=0,1,2,...,
m_0 \equiv m(\pi^{(0)})$,
\be
G^{(MM)} (k) =- \sum^\infty_{n=0}
\frac{(c_n^{(M)})^2}{k^2+m^2_n},~~ G^{(M)} (k) =-\sum^\infty_{n=0}
\frac{c_nc_n^{(M)}}{k^2+m^2_n}.\label{8}\ee

In the next section we shall determine the coefficients $c_n,
c_n^{(M)}$ and $m_n$ for the  lowest states, and now we define
$\lan \bar q q\ran$ and $f_\pi$ in terms of spectral sums
(\ref{8}). From (\ref{6}) and (\ref{7}) one has
\be
\lan \bar \psi \psi\ran_M =-2N_c
\sum^\infty_{n=0}\frac{c_nc_n^{(M)}}{m^2_n}, ~~f_\pi^2 =2
N_c\sum^\infty_{n=0}\frac{(c_n^{(M)})^2}{m^4_n}.\label{9}\ee

The coefficients $c_n$ and $c_n^{(M)}$ differ by the presence of
the vertex operator $M_S\equiv M(0)$  in the latter which is a
constant computed in Appendix 1, therefore one has
$c^{(M)}_n=M(0)c_n$, and limiting oneself to the first term in the
sum (\ref{9}) one obtains
\be
|\lan \bar \psi \psi\ran_M|\geq m^2_0f^2_\pi
\frac{c_n}{c_n^{(M)}}= \frac{m^2_0f^2_\pi}{M(0)}.\label{10}\ee

Inserting $M(0)=148$ MeV from Appendix 1 and $|\bar\psi\psi|=(225$
MeV$)^3$, $f_\pi=94$ MeV, one obtains $m_0\cong 437$ MeV which is
close to the value $m_0=400$ MeV calculated in Appendix 3. On the
other hand, the sum (\ref{9}) for $\lan \bar \psi \psi\ran$ is
converging more slowly than that for $f_\pi$, and  therefore  one
has inequality in (\ref{10}) due to the  presence of higher terms
in $\lan \bar \psi \psi\ran$ .

\section{Calculation of $\lan \bar q q\ran$ and $f_\pi$}

The integration region in the space-time integrals in (\ref{3}),
(\ref{6}) can be split in two parts: $|x-y|> T_g$ and $|x-y|\leq
T_g.$ In the first (long distance) region the relativistic local
potential-type dynamics  sets in at space-time  distances
exceeding $T_g$ \cite{15}, \cite{16} and the result can be
expressed in terms of the spectrum, as will be done below in this
section. The second region can be treated in the OPE formalism
\cite{17} and is considered in the Appendix 4. It is shown there
that the contribution of this region is parametrically small in
the parameter $\sigma T^2_g\ll 1$.  Only  the long-distance
contribution is calculated in this section below.

 We
start with the calculation of $f_\pi$ and to this end we write the
$q\bar q$ Green's function $G^{(MM)}(k)$ in terms of c.m. and
relative coordinates  as follows (another derivation is given in
Appendix 2): $$ - G^{(MM)}(k) =\int d^4 X G^{(MM)}(\ver_{12}=0;
\veR =0, \ver'_{12}=0, \veR'=\veX, T)e^{i\vek\veX}= $$ $$=
M^2(0)\sum_n |\varphi_n(0)|^2\frac{d^3\veP}{(2\pi)^3} dTd^3\veX
e^{-E(P)T-i\veP\veX+i\vek\veX}=$$ \be = M^2(0)\sum_n
|\varphi_n(0)|^2\int^\infty_0 e^{-E(\vek)T}dT=M^2(0)\sum_n
\frac{|\varphi_n(0)|^2}{\sqrt{m^2_n+\vek^2}}.\label{11}\ee

Expanding  (\ref{11})  in $\vek^2$ and comparing to (\ref{7}) one
finds $f^2_\pi$,
\be
f^2_\pi =N_cM^2(0)\sum^\infty_{n=0}
\frac{|\varphi_n(0)|^2}{m_n^3}.\label{12}\ee Comparing (\ref{9})
and (\ref{12}) one finds
\be
c_n^{(M)} = \sqrt{\frac{m_n}{2}} M(0) \varphi_n(0),
c_n=\sqrt{\frac{m_n}{2}}\varphi_n(0).\label{13}\ee

In a similar way one computes $\lan \bar q  q\ran$ from (\ref{6})
and finds
\be
-\lan \bar q q\ran = N_cM(0)\sum^\infty_{n=0}
\frac{|\varphi_n(0)|^2}{m_n}.\label{14}\ee

Here $\varphi_n(\ver)$ is the 3d  spin-singlet wave-function of
$q\bar q$ system, as obtained e.g. in the relativistic Hamiltonian
method of FCM \cite{15}, or else in the Bethe-Salpeter equation
with the kernel not depending on relative time, as it  is
discussed in Appendix 2.

The accuracy of the method with  respect  to calculation of
$|\varphi_n(0)|^2$ can be checked by comparison  of predicted
leptonic width with experiment, as it is done in Appendix 3.
Taking into account both color Coulomb and confining interaction
one has
\be
|\varphi_n(0)|^2 =\frac{\mu_n(\sigma+\frac43\alpha_s\lan
\frac{1}{r^2}\ran)}{4\pi}\label{15}\ee where $\mu_n$ is the
constituent energy (mass) computed through $\sigma$ \cite{15}; we
refer the reader to the Appendix 3 for the details of calculation
of $m_n$ and $|\varphi_n(0)|^2$.

As it is shown in \cite{16} and discussed in Appendix 2 , the
masses $m_n^2, \mu_n^2$ grow linerly with $n$ in the large $N_c$
limit, hence the sum (\ref{14})    for $\lan \bar q q\ran$ is
formally diverging if the spectrum of radially excited mesons
extends to infinitely large masses. In fact the experimental
spectrum can be followed up to the mass values around
$m_{cont}\cong 2.5$ GeV, where resonances become very wide and
strongly mix between themselves and with hybrids, forming the
continuum of states. Following the ideology of the QCD sum rules
\cite{17} one could replace this continuum by the perturbative
diagrams, which  do not contribute to  $\lan \bar q q\ran$.
Therefore we shall keep the first 3 terms in the sum (\ref{14})
over $n$ (the term with $n=3$ gives negligible contribution). As
was mentioned in the begining of this section the relativistic
potential description of $ G^{(MM)} (\ver_{12}, \veR, 0, \ver_{12},
\veR', T)$ is possible only for the time $T\ga T_0$, while for
$T<T_0, T_0\sim T_g$ one should use the properties of the $\bar q
q$ Green's function $\tilde G,$ as given by the OPE \cite{17}. As
it is discussed in Appendix 4, the region of small times and
relative distances covered by the OPE treatment, gives a
contribution to $\lan \bar q q\ran$ proportional to
$O(\sigma^{5/2} \frac{T_0^4}{T_g^2},\sigma m) $ and therefore can
be disregarded for light quarks and small $ T_0<T_g$. As a result
one should exclude from the integration over $dT$ in (\ref{11})
the region $(0,T_0)$ which brings about the following factor in
(\ref{14}) instead of $1/m_n$
\be
\frac{1}{m_n} \to \frac{e^{-m_nT_0}}{m_n}\label{16}\ee and in
(\ref{12})
\be
\frac{1}{m^3_n} \to \frac{e^{-m_nT_0}}{m^3_n}
(1+m_nT_0).\label{17} \ee

Keeping for $\lan \bar q q\ran$ the first 3 terms in the sum
(\ref{14}) and 2 terms in (\ref{12}) one has
\be
-\lan \bar q q\ran= N_cM(0)\left \{ \frac{\varphi_0^2(0)
e^{-m_0T_0}}{m_0} + \frac{\varphi_1^2(0) e^{-m_1T_0}}{m_1} +
 \frac{\varphi_2^2(0) e^{-m_2T_0}}{m_2}\right\}
\label{18} \ee
\be
f^2_\pi= N_cM(0)\left \{ \frac{\varphi_0^2(0)
e^{-m_0T_0}}{m_0^3}(1+m_0T_0)+ \frac{\varphi_1^2(0)
e^{-m_1T_0}}{m_1^3} (1+m_1T_0) \right\}.
 \label{19} \ee

Using (\ref{A.14}), (\ref{A.19}) one has
\be
\varphi_0^2(0) =\frac{0.109 {\rm{GeV}}^3}{4\pi},~~ \varphi_1^2(0)
=\frac{0.097 {\rm{GeV}}^3}{4\pi},~~ \varphi_2^2(0) =\frac{0.115
{\rm{GeV}}^3}{4\pi}.\label{20} \ee $$ m_0=0.4{\rm{GeV}},~~ m_1
=1.35 {\rm{GeV}},~~ m_2= 1.85 {\rm{GeV}}.$$

For a reasonable estimate we put  $T_0=T_g=1{\rm{GeV}}^{-1}$ and
the value $M(0) =0.148$ GeV from Appendix 1, and obtain.
\be
-\lan \bar q q\ran = (0.195{\rm{GeV}})^3,~~ f_\pi =0.094
{\rm{GeV}}.\label{21}\ee

 One can
 check  that the behaviour of (\ref{18}) for $\lan\bar q q \ran$  at
small $T_0$ is smooth, e.g. when changing $T_0$ from $T_g=1$
GeV$^{-1}$ to 0.5 $T_g$, the result changes by roughly 10\%.

 To check the sensitivity to the change of $T_g$, we
have taken $T_g = 1/1.5$ GeV$^{-1}$ and recalculated all
quantities, e.g. from (\ref{A.7}) one has $M(0)=0.12$ GeV. The
resulting values are not much changed from (\ref{21}),
\be
-\lan \bar q q\ran (T_g =\frac{1}{1.5}{\rm{GeV}^{-1}}) = (0.189
{\rm{GeV}})^3, ~ f_\pi =0.076 {\rm{GeV}}.\label{22}\ee

It is remarkable that $f_\pi$ in (\ref{21}) is very close to the
value   obtained from the pion decay and used in the chiral
perturbation theory \cite{18} $f_\pi =93$ MeV. At the same time
$|\lan \bar q q\ran |$ is somewhat less than the standard value
(240 MeV)$^3$,  and we discuss in the concluding section section
 the scale dependence and comparison to existing lattice measurements.

\section{Discussion and conclusions}

The quark condensate and $f_\pi$ are given by Eqs.(\ref{18}),
(\ref{19}) and  (\ref{22}), where all quantities can be expresses
through $m, \sigma$ and $T_g$, since $\varphi^2_n(0)$, $m_n$ and
$M(0)$ are expressed through these quantities, while $T_0$ can be
taken in the region of plateau and e.g. equal to $T_g$. In this
way one obtains $(m=0, \sigma=0.18 $ GeV$^2,~~ T_0 =T_g=1$
GeV$^{-1}$, and Eq.(\ref{22}) for $\lan \bar q q\ran$)
\be
f_\pi\cong 0.094 {\rm GeV} ,~~-\lan \bar q q\ran \cong (0.20 {\rm
GeV})^3.\label{25}\ee Several corrections should be added to this
results.
 First of all, the short  distance contribution to $\lan \bar q
 q\ran$ is of relative order $\sqrt{\sigma} T_g\sim 0.45$ and can
 substantially increase the result. Another essential point is
  the value of $T_g$, which increases
  in the presence of dynamical quarks, and can be smaller if  gluelump data \cite{19}
are instead taken into account,  $T_g=0.7$ GeV$^{-1}$. This
influences significantly the value of $M(0)$, however an
independent check can be made since $M(0)$ also enters the strong
decay matrix element, and the value $M(0)= 0.148$ GeV is
reasonably close to the phenomenological value known from the
$^3P_0$ model  \cite{20}.

We are now in position to compare (\ref{25}) with the lattice
data. There the computation was  done  in the quenched case for
Wilson fermions \cite{21} and also for the overlap action
\cite{22}. Before using  the evaluation coefficient for $\lan \bar
q q\ran$, one can compare the result (\ref{25}) which does not
contain any scale $\mu$, and any evolution corrections, with the
so-called Renormalization  Group Invariant (RGI) lattice
measurements, which yield \cite{21} \be -\lan \bar q
q\ran^{RGI}_{lat}= [(206\pm  44 \pm 8 \pm \pm 5) {\rm~
MeV}]^3.\label{26}\ee This value is in reasonable agreement with
(\ref{25}).  As the next step we take  the evolution coefficient
for  $\lan \bar q q\ran $ computed in  \cite{23} $(n_f =0, N_c=3)$
\be
 C_s^{\overline{MS}} (\mu) = [\alpha_s (\mu) ]^{-4/11} \{ 1- 0.219
 \alpha_s - 0.1054 \alpha_s^2\}.\label{27}\ee
 For $\mu=2$ GeV taking $\alpha_s\approx 0.3 $,
 and identifying $\lan \bar q q\ran$ in (\ref{25}) with $\lan \bar q q\ran^{RGI}$
  one obtains for the
 long-distance contribution to the condensate
 \be
 \lan \bar q q\ran (\mu =2{\rm GeV}) \cong \lan \bar q q\ran^{RGI}
 C_s^{ \overline{MS}}\cong -(225 {\rm MeV})^3.\label{28}\ee

 This value, given in the abstract of the paper, is obtained
 without  inclusion  of the coefficient used on the lattice
 \cite{21} to calculate the transition  from the lattice  RGI result  to the $ \overline{MS
 }$ scheme, this coefficient is anyhow close to unity.

 The lattice value at $\mu=2$ GeV for Wilson quarks in \cite{20}
 \be
 \lan \bar q q \ran^{ \overline{MS}}(\mu=2 {\rm  GeV}) = -
 [(242\pm  9 ) {\rm MeV})]^3\label{29}\ee
  and differs from the result \cite{22}: -- (282 (6)
  MeV)$^3\left(\frac{a^{-1}}{1766 {\rm MeV}}\right)^3$.
   An independent estimate from the QCD sum rules yields \cite{24}
   \be
   \lan \bar q q\ran (\mu=M_N) =- [(225\pm 9) {\rm
   MeV}]^3.\label{30}\ee
   As a result one can see that our long-distance contribution to $\lan \bar q  q\ran $, Eq. (\ref{28}),
   is somewhat smaller than the lattice
   data (\ref{29}), but is certainly in the same ballpark, and the evaluation of the short-distance contribution
    is
   important to improve the accuracy of calculation.

   At the same time the resulting value $f_\pi$ (\ref{25}) is in
   good agreement with the standard value, obtained from the pion
   decay and used in the chiral perturbation theory \cite{18}.

   The method used above can be easily applied to the case of
   nonzero quark mass $m$ and the SU(3) flavour group to calculate
   $\lan \bar s  s\ran$, $f_K$ etc., which will
   be published elsewhere \cite{25}.

   The  financial support of INTAS grants 00-110 and 00-366 is
   gratefully acknowledged.

\newpage

 \setcounter{equation}{0}
\renewcommand{\theequation}{A.\arabic{equation}}

\begin{center}
{\bf Appendix 1 }\\

 \vspace{0.5cm}

{\large Calculation of the vertex mass $M(0)$}
\end{center}

One starts with the definition of the nonlocal mass operator
$M_S(u,v)$, given in \cite{9,10} (see e.g. Eq.(24) in \cite{10})
\be
M_S (u,v) = (\gamma_\mu \Lambda (u,v) \gamma_\mu) _{sc}
J(u,v) .\label{A.1}\ee

The mass operator enters in the gauge-invariant Green's functions,
see e.g. Eq.(3), via the quark propagator $\Lambda (x,y)$, Eq.
(4), where $M_S(z,u)$  enters at all intermediate points, and also
at initial and final points $\frac{x+\bar x}{2}$ and $\frac{y+\bar
y}{2}$, where the nonlocal pion $\phi(x, \bar x)$ is emitted.
According to the prescription given in \cite{10}, we choose the
set of contours $C(z)$ for all intermediate points $z$ in the
Green's function $G(\frac{x+\bar x}{2}, \frac{y+\bar y}{2})$,
which minimizes the mass eigenvalues. One simple choice is to take
the contours $C(z)$ from $z$  a long the shortest way to the $x_4$
axis passing through $\frac{x+\bar x}{2}$ and $\frac{y+\bar
y}{2}$, and along $x_4$ axis to the origin  at the point
$\frac{x+\bar x}{2}$.

 When $M_S $ is situated at the initial or final
point of the $q\bar q$ Green's function, i.e. at the points
$M_S(x,\bar x)$ or  $M_S(y,\bar y)$, where the $q\bar
q$ or pion  is created or annihilated, then it is convenient to
choose  points $x,\bar x$ on the axis 1 with the origin at
$\frac{x+\bar x}{2}$.
  In this way one obtains for $x_1> 0,\bar x_1\equiv y_1<0$,
  $y_4\equiv\bar x_4$.
\be
J(x, y) = \int^{x_1}_0 du \int_{y_1}^0 dv D(u-v,
x_4-y_4).\label{A.2}\ee

It is convenient to use for $D$ the Gaussian form, \be D(\vex,
x_4) = D(0) e^{-\frac{{\tiny \vex}^2+  x^2_4}{4Tg^2}} =
\frac{\sigma}{ 2\pi T^2_g} e^{-\frac{{\tiny \vex}^2+ x^2_4}{ 4
T^2_g}}, \label{A.3}\ee which yields \be J(x,y)
=\frac{\sigma}{\pi} e^{-\frac{(x_4-y_4)^2}{4T^2_g}}
(1-e^{-\frac{({\tiny \vex-\vey})^2}{4T^2_g}}).\label{A.4}\ee

Now one has to estimate the scalar part of the  quark Green's
function $\Lambda(x,y)$ in (\ref{A.1}), for which in \cite{8} it
was found that it behaves as a smeared $\delta$- function with the
smearing radius equal to $1/\sqrt{\sigma}$ (see Eq.(24) in the
second ref. of \cite{8}). We simplify this  form taking
\be
\Lambda(x,y) = \left( \frac{\sigma}{\pi}\right)^{3/2} e^{-({\tiny
\vex -\vey})^2\sigma};~~ \int \Lambda (x,y) d^3 (\vex -\vey) =1.
\label{A.5}\ee To obtain the localized form of the vertex function
\be
M_S(x) \equiv M(0) =\int M^{(0)}_S(x,y)  d^4 (x-y)
\label{A.6}\ee we substitute in (\ref{A.6}) $J(x,y)$ from
(\ref{A.4}) and $\Lambda(x,y)$ from (\ref{A.5}) to get finally.
\be
M(0) = \left [ 1- \left ( \frac{\sigma^4 T^2_g}{\sigma^4 T_g^2+1}
\right)^{3/2} \right] \frac{2\sigma T_g}{\sqrt{\pi}} \equiv \eta
\frac{2\sigma T_g}{\sqrt{\pi}}. \label{A.7}\ee In the limit
$\sigma T^2_g\to 0 $ one obtains $M(0)\approx \frac{2\sigma
T_g}{\sqrt{\pi}}$, i.e. exactly the value which appears in the
strong decay vertex of the string in \cite{20}. This is not
surprising since in both cases $M(0)$ is a mass corresponding to a
piece  of the  string  with the length of the order of $T_g$,
hence $M(0) \sim \sigma T_g$.  The factor $\eta$ in (\ref{A.7})
describes the attenuation due to the nonlocality of
$\Lambda(x,y))$ at small $|\vex - \vey |$ for light quarks. For
heavy quarks this factor tends to zero since the localization of
$\Lambda
 (x,y) $ becomes more strong, indeed the quark Green's
function $\Lambda$ for $m\to \infty$ is proportional to
$\delta^{(3)} (\vex-\vey)$ see \cite{8}. Effectively  for nonzero
$m$ this can be described replacing $\eta $ in (\ref{A.7}) by the
factor \be \eta \to \eta (m)=\left [ 1- \left ( \frac{(\sigma
T_g+m)4T_g}{(\sigma T_g +m) 4 T_g+1}\right )^{3/2}
\right].\label{A.8}\ee

For light quarks and $\sigma =0.18 $ GeV$^2,~~ T_g=1$ GeV$^{-1}$,
the  factor $\eta\equiv  \eta(0)$ is $1-0.27 \cong 0.73,$  and
from (\ref{A.7}) one gets $M(0)\approx 0.148$ GeV.

\begin{center}
{\bf Appendix 2 }\\

 \vspace{0.5cm}

{\large Derivation of the spectral representations, Eqs.
(\ref{12}) and (\ref{13})}
\end{center}

Consider the $q\bar q$ Green's function of the type given in Eq.
(\ref{6})
\be
G_{\Gamma} (x,y) = \lan tr \Gamma \Lambda (x,y) \Gamma \Lambda
(y,x)\ran\label{A2.1}\ee where $\Gamma=\gamma_5, \gamma_\mu,...$,
and the $q\bar q$ Green's function in the $4\times  4$ spinor
representation,
\be
G^{(q\bar q})_{\alpha\beta, \gamma\delta} (x,\bar x; y,\bar
y)=\lan \Lambda_{\alpha\beta} (x,y) \Lambda_{\gamma\delta} (\bar
y, \bar x)\ran.\label{A2.2}\ee

Following the standard procedure from \cite{26} one can introduce
the c.m. and relative coordinates, e.g.
\be
X=\frac{x+\bar x}{2}, Y=\frac{y+\bar y}{2}, r=x-\bar x, r'=y-\bar
y \label{A2.3}\ee and define
\be
G^{(q\bar q)} (x,\bar x; y, \bar y)= \int d^4 Pe^{iP(X-Y)}
\frac{d\varepsilon}{2\pi} e^{-i\varepsilon r_0 } G_P (\ver, \ver',
\varepsilon, \ver'_0)\label{A2.4}\ee $G_P$ satisfies an equation
\be
(E-E_2-H_1) (E_2-H_2) G_P=\beta_1\beta_2\hat 1\label{A2.5}\ee
where $E=E_1+E_2= P_0, E_1-E_2= 2\varepsilon$, and
\be
H_i=m_i\beta_i + \vep \veal_i + \beta_i M_S.\label{A2.6} \ee

At this point one can exploit the property of $H_i$ that  it does
not depend on relative time $r_0$, and therefore one can integrate
in (\ref{A2.4}) over    $d\varepsilon$ with the result
\cite{26,27}
\be
G_P\left|_{\begin{array}{l} {r_0=0}\\ {r'_0=0}
\end{array}}\right. = \beta_1 \beta_2 \int^\infty_{-\infty}
\frac{d\varepsilon/2\pi}{(E_1-H_1)(E_2-H_2)} = i\beta_1\beta_2
\frac{1}{E-{\hat H}} , \hat H \equiv H_1+H_2.\label{A2.7} \ee

As the result one obtains
\be
\int d^4 (X-Y) G^{(q\bar q)} (x,\bar x; y, \bar y)
\left|_{r_0=r'_0=0}\right. = \lan \ver \left|
\frac{i\beta_1\beta_2}{\hat H} \right| \ver'\ran = \sum_n \lan
|\ver n\ran \frac{i\beta_1\beta_2}{E_n} \lan n |
\ver'\ran.\label{A2.8}\ee One can now express $G_\Gamma$, with
$\Gamma=\gamma_5$,
\be
\int d^4 (X-Y) G_\Gamma (x,y) =i \sum^\infty_{n=0}
\frac{\psi_n(0)\psi_n^+(0)}{E_n}\label{A2.9}\ee where we have
defined the relativistic wave-function $\psi_n(\ver )\sim \gamma_5
\lan \ver|n\ran$, $\psi^+_n (\ver) \sim \beta_1\beta_2\lan
n|\ver\ran$, satisfying the Hamiltonian equation
\be
\hat H\psi_n (\ver) =E_n \psi_n (\ver) . \label{A2.10} \ee As it
is known from dynamical calculations with the Bethe-Salpeter
equation \cite{28} with the scalar confining kernel the dominant
role in $\psi_n(\ver)$ is played by the $^1S_0$ component
$\varphi_n(\ver)$ of the wave-function, which satisfies the
relativistic Schroedinger equation with the hyperfine interaction,
discussed in the Appendix 3. Therefore one can identify
$\psi_n(\ver) \to \varphi_n(\ver), E_n\to m_n$ and the Eq.
(\ref{6}) with the help of (\ref{A2.9}) goes over into Eq.
(\ref{14}).

\begin{center}
{\bf Appendix 3 }\\

 \vspace{0.5cm}

{\large Calculation of the  masses $m_n$ and $\varphi_n(0)$
through $\sigma$ in FCM}
\end{center}

The mass eigenvalue $\bar m_n$ of the spin-averaged  state
$\frac{3 m_n^{\rho)} + m_n^{(\pi)}}{4}$ for $L=0$ can be written
as \cite{15,16} \be \bar m_n=M_0(n,0) +\Delta_{SE}
+\Delta_C,\label{A.9} \ee where $M_0(n,0)$ is the eigenvalue of
the spinless Salpeter equation, which can be written as
\be
M_0(n,0)=4 \mu_0 (n)= 4\lan
\sqrt{\vep^2+m^2}\ran_{n0}.\label{A.10} \ee

For $m=0$, $\mu_0(n,0)$ is expressed through $\sqrt{\sigma}$ and
dimensionless coefficients $a(n)$ -- zeros of Airy functions
\cite{15}
\be
\mu_0(n) = \sqrt{\sigma} \left ( \frac{a(n)}{3}\right)^{3/4},~~
a(0) = 2.338, ~~ a(1)=4.088. \label{A.11a} \ee

Taking into account nonzero $m$ one finds $\mu_0(n)$ from the
equation \be 1=\frac{m^2}{\mu^2_0} +
\frac{\sigma^{2/3}}{3\mu_0^{4/3}} a(n). \label{A.11} \ee For large
$m\gg \sqrt{\sigma}$ the solution of (\ref{A.11}) is
\be
\mu_0^2(n) \cong m^2 \left[1+ \frac{a(n)}{3} \left(
\frac{2\sigma\mu}{m^2(m+\mu)}\right)^{2/3}\right]. \label{A.12}\ee
The term $\Delta_{SE}$ is the self-energy correction \cite{19}
which can be written as
\be
\Delta_{SE} (n) =- \frac{4\sigma\eta (m)}{\pi\mu_0(n)}
\label{A.13} \ee and $\eta(m) $ is computed through $m, $ for
$m=0$ it is $\eta(m=0) =0.9 \div 1.$

Taking all contributions into account one obtains for the light
quarks $(m=0)$
\be
\bar m_0= 0.652 {\rm{GeV}},~~ \bar m^2_n=\bar m^2_0 +\Omega_0 n,
n=0,1,2..\label{A.14}\ee where $\Omega_0$ is computed solely
through $\sigma$ and is equal $\Omega_0\cong 1.6 $ GeV$^2$, which
is close to the experimental slope $\Omega_{exp} (L=0) =1.64\pm
0.11 $ GeV$^2$, see \cite{16} for refs. and discussion.

Now we take into account the hyperfine interaction which produces
the HF splitting between $\rho$ and $\pi$ states.

\be
\Delta_{HF} =\Delta_{HF}^{Pert} +\Delta_{HF}^{NP},~~
\Delta_{HF}^{Pert} =
\frac{8\alpha_s(\mu_{HF})|R_n(0)|^2}{9\mu_0^2(n)}.\label{A.15}\ee
Here $R_n(0)=\sqrt{4\pi} \varphi_n (0)$ is the radial meson w.f.,
which can be found also from the leptonic width of $\rho$ meson.
One has
\be
|R_0(0)|^2=\mu_0(0) (\sigma +\frac{4}{3} \alpha_s \lan r^{-2}\ran)
=
\left\{ \begin{array} {ll} 0.091 {\rm{GeV}}^3, & \alpha_s=0\\
0.109{\rm{GeV}}^3, & \alpha_s=0.3.\end{array}
\right.\label{A.16}\ee

These values can be checked $vs$  the leptonic width of $\rho$,
$\Gamma_{e^+e^-}=\left\{ \begin{array}{l} 6.36
{\rm{KeV}},\alpha_s=0\\7.62{\rm{KeV}},\alpha_s=0.3\end{array}\right.
$, while $\Gamma^{\exp}_{e^+e^-} = (6.85\pm 0.11){\rm{KeV}}.$

Thus one obtains $\Delta^{Pert}_{HF}$ from (\ref{A.15}),
$\Delta^{Pert}_{HF} = \left\{
\begin{array}{l}
0.26{\rm {GeV}}, \alpha_s=0\\0.24{\rm
{GeV}},\alpha_s=0.3.\end{array}\right.$

The nonperturbative part $\Delta^{NP}_{HF}$ is expressed through
the correlator $D(x) $ \cite{20} and depends on the accepted value
of $G_2 \equiv \frac{\alpha_s}{\pi} \lan F^a_{\mu\mu}
F^a_{\mu\nu}\ran$,
\be
\Delta_{HF}^{NP} \cong 50 MeV \left(
\frac{G_2}{0.012{\rm{GeV}}^4}\right).\label{A.17}\ee We take two
values of $G_2=G^{st}_2 =0.012$ GeV$^4$ and $G_2=2G^{st}_2$. Thus
one obtains  for the lowest mass of PS state in the $q\bar q$
approach (no chiral effects)
\be
m_0 = \bar m_0 -\frac34 \Delta_{HF} = [0.652 -\frac34 (0.3\div
0.35)]{\rm{GeV}}=(0.39-0.43){\rm{GeV}}.\label{A.18}\ee

 As a result  we accept the following values for $m_0$ and $m_1$
(the latter is calculated in the same way using (\ref{A.14}) and
$\Delta_{HF}(n=1))$
\be
m_0=0.4 {\rm{GeV}},~~ m_1=1.35 {\rm{GeV}}.\label{A.19}\ee

\begin{center}
{\bf Appendix 4 }\\

 \vspace{0.5cm}

{\large Small distance contribution to $\lan \bar q q
\ran$}
\end{center}

To separate the small-distance contribution we start from Eq.
(\ref{6}) where we take into account the nonlocal structure of
$M_S(u,v)$ and put $m=0$,
\be
tr \Lambda_{xx} =\int d^4 u d^4 y tr (\gamma_5 \Lambda (x,y)
\gamma_5 M_S (y,u) \Lambda (u,x) ).\label{A.20} \ee In the limit
when one keeps the most singular part -- the free part of
$\Lambda(x,y)$, one has
\be
\Lambda(x,y) \approx \Lambda_{free} (x,y) = \frac{1}{2\pi^2}
\left( \frac{\hat x-\hat y}{(x-y)^4} +\frac{m}{2(x-y)^2} \right)
+...\label{A.21}\ee

From (\ref{A.1}) one can derive the behaviour of $M_S(y,u)$ at
small $|y-u| \leq T_0 , T_g$,
\be
M_S(y,u) \sim \frac{\sigma}{T_g^2} c|y-u|^2 \Lambda(x,y)
,\label{A.22}\ee where the coefficient $c$ is of the order of
unity.

 The nonperturbative part of $\Lambda(x,y)$ is not singular
(modulo logarithms) and proportional to $\sigma^{3/2}$, c.f.
Eq.(A.5) (apart from the OPE part of $\Lambda$ which has $m\lan
\bar q q\ran$ and $\lan F^2\ran$  terms and even less singular at
small $x$). Finally inserting (\ref{A.22}), (\ref{A.21}) into
(\ref{A.20}) and integrating in the region $|x-y|, |y-u|,
|u-x|\leq T_0$ one can write the short-distance contribution to
(\ref{A.20}) as
\be
tr\Lambda_{xx} ({\rm{small~~ distance}}\leq T_g)=O(\sigma m,
(\sigma^{5/2} \frac{T_0^4}{T^2_g}).\label{A.23}\ee As it is seen
from (\ref{18}) the long distance part is $O(\sigma^2T_g)$ and is
dominant at $\sigma T^2_g\to 0, T_0\leq T_g$.

\end{document}